\documentclass[12pt,report]{elsart}

\usepackage{hyperref}
\usepackage{epsfig,graphics,colordvi}
\usepackage{amsmath}
\usepackage{amssymb}

\begin{document}
\begin{frontmatter}
\Large{\title{On causality, unitarity \\ and perturbative expansions }}
\author[GSI,ITEP]{I.V. Danilkin,}
\author[GSI,ITEP]{A.M. Gasparyan,}
\author[GSI]{and M.F.M. Lutz}
\address[GSI]{Gesellschaft f\"ur Schwerionenforschung (GSI),\\
Planck Str. 1, 64291 Darmstadt, Germany}
\address[ITEP]{Institute for Theoretical and Experimental Physics,\\
117259, B. Cheremushkinskaya 25, Moscow, Russia}
\begin{abstract}

We present a pedagogical case study how to combine micro-causality and unitarity based on a perturbative approach.
The method we advocate constructs an analytic extrapolation of partial-wave scattering amplitudes that is
constrained by the unitarity condition. Suitably constructed conformal mappings help to arrive at a systematic
approximation of the scattering amplitude. The technique is illustrated at hand of a Yukawa interaction.
The typical case of a superposition of strong short-range and weak long-range forces is investigated.

\end{abstract}
\end{frontmatter}

\section{Introduction}

Recently a novel scheme for studying hadronic interactions beyond the threshold
region was introduced by two of the authors \cite{Gasparyan:2010xz}.  The main objective
of that work is a controlled realization of the causality and unitarity condition
in a perturbative application of the chiral Lagrangian. The starting point are partial-wave dispersion relations.
A generalized potential is constructed from the chiral Lagrangian in the subthreshold region and analytically
extrapolated to higher energies. The partial-wave scattering amplitudes are obtained as solutions of
non-linear integral equations.

The purpose of the present letter is an illustration of the method for a schematic system where the
exact solution is known. This will shed further light on its usefulness. We consider non-relativistic
Yukawa interactions of various strengths and ranges. Though a quantitative description of for instance
the pion-nucleon or nucleon-nucleon system is not justified in terms of static Yukawa potentials there are
essential features grasped by a Yukawa ansatz. The long-range part of the nucleon-nucleon force can be understood
in terms of a static Yukawa potential derived from the pion-exchange process. The important feature is a
left-hand cut in the partial-wave scattering amplitudes very close to the physical region. In this respect the
pion-nucleon system is similar, as its long-range force generated by the u-channel nucleon exchange process defines
close-by left hand cuts as well.

\section{Expansions with a Yukawa potential}
\label{Yukawa_single}

We consider the quantum mechanics of two particles with masses $m_1$ and $m_2$
defined by a single Yukawa potential. The system is characterized by the reduced mass parameter
$m =m_1 \,m_2\,/(m_1 +m_2)$. The scattering process in a given partial wave with angular momentum $l$
is determined by the  $t$-matrix, satisfying
the nonrelativistic partial-wave Lippmann-Schwinger equation
\cite{Brown:1976}:
\begin{eqnarray}
&& \langle k'|\,t_l(q^2)|k\rangle =\langle k'|V_l\,|k\rangle
+\frac{4 \,m}{\pi}\int_0^\infty k''^2\,dk''\,\frac{\langle k'|V_l\,|k''\rangle\, \langle k''|\,t_l(q^2)|k\rangle}
{q^2-k''^2+i\,\epsilon}\,,
\label{LSPW}
\end{eqnarray}
with $q,k,k'$ being the center of mass on-shell and initial and final off-shell momenta respectively.
The Yukawa potential projected onto angular momentum reads
\begin{eqnarray}
&&\langle k'|V_l\,|k\rangle
=\frac{g}{2\,k' k}\,Q_l\left(\frac{k^2+k'^2+ \mu^2}{2\,k' k}\right) \,, \qquad Q_0 (x) = \frac{1}{2}\,\log \frac{x+1}{x-1}\,,
\nonumber\\
&& Q_l (x) = \frac{1}{2}\,P_l(x) \,\log \frac{x+1}{x-1} - \sum_{k=1}^l \,\frac{1}{k}\,P_{k-1}(x)\,P_{l-k}(x)\,,
\label{Yukawa}
\end{eqnarray}
with the conventional Legendre polynomials $P_l(x)$.
The Yukawa mass $\mu$ characterizes the range of the interaction. For simplicity and definiteness
we assume nonrelativistic kinematics.

\begin{figure}[t]
\begin{center}
\parbox[c]{7.5cm}{\includegraphics*[width=6cm,height=6cm]{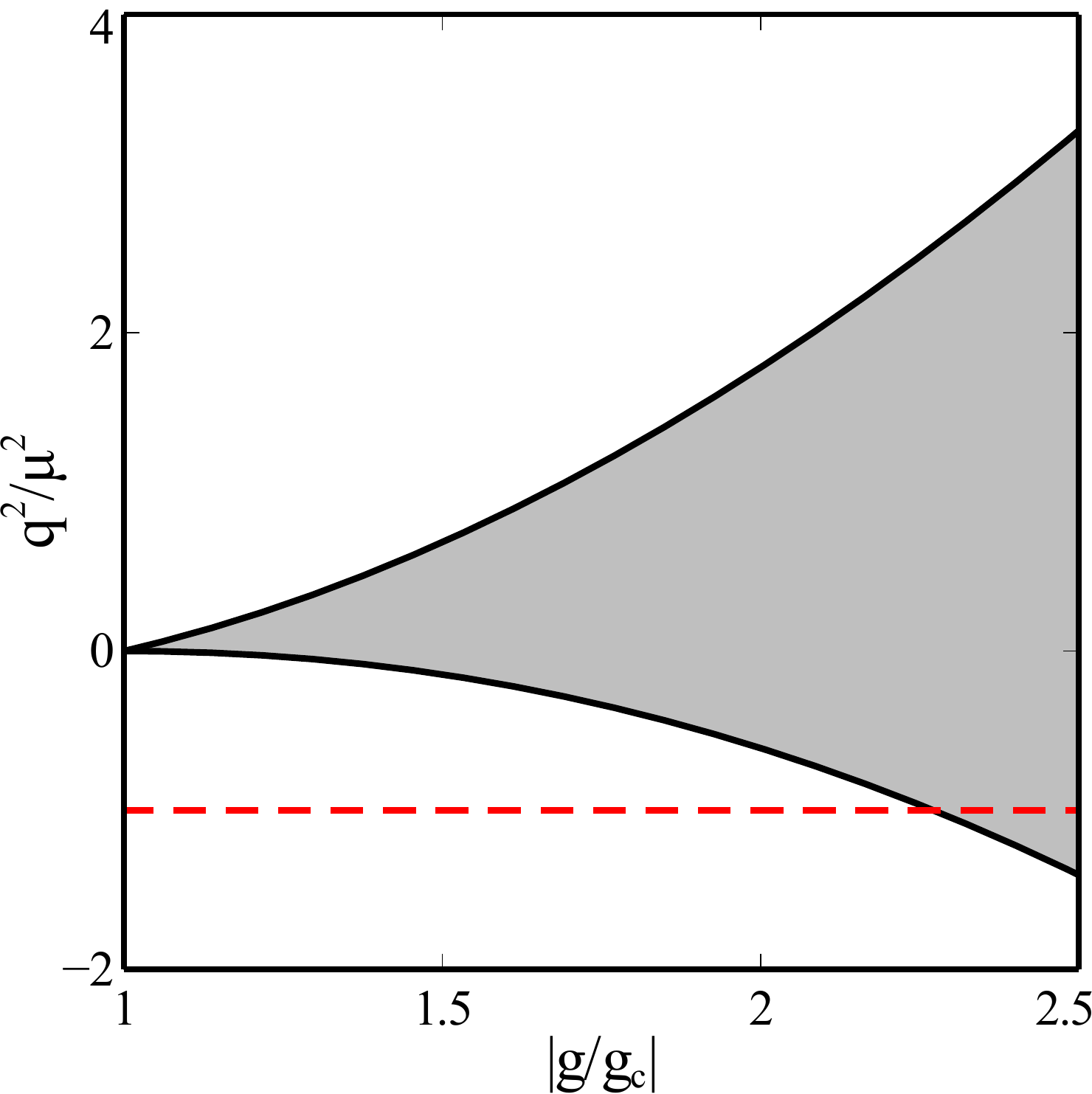}}
\parbox[c]{5.0cm}{\caption{Convergence domain of the Born series for the $s$-wave
Lippmann-Schwinger equation (\ref{LSPW}) with (\ref{Yukawa})  as a function of the coupling
strength $g/g_c$.
The dashed curve locates the branch point caused by
the 2nd order Born term.}}
\end{center}
\label{fig:convergence}
\end{figure}

The integral equation (\ref{LSPW}) can be expanded
in powers of the coupling constant $g$, defining the Born series. The expansion
converges for certain values of $g$ and $q^2$ \cite{Weinberg:1963zz,Lutz:1999yr}.
The area of convergence in the plane of $q^2/\mu^2$ and $|g/g_c|$ is shown
in Fig. 1 for the $s$-wave potential.
For coupling constants, $|g| < g_c $,  smaller than the critical one
\begin{eqnarray}
g_c \simeq  1.68 \,\frac{\mu}{2\,m} \,,
\label{}
\end{eqnarray}
the whole region $-\infty < q^2<+\infty$ is perturbative. Within the shaded area the Born series does not converge.
If the coupling constant exceeds its critical value
the scattering amplitude is perturbative only either for sufficiently small or large $q^2$ as shown in
Fig. \ref{fig:convergence}.
In the case of attraction with $g<0$ the first bound state is formed at $g=-g_c$ \cite{Ericson:1988gk}.
An important observation is that even if perturbation theory fails to converge close to threshold,
it does converge for sufficiently large negative $q^2$.

\begin{figure}[t]
\centerline{  \includegraphics*[width=6cm,height=6cm]{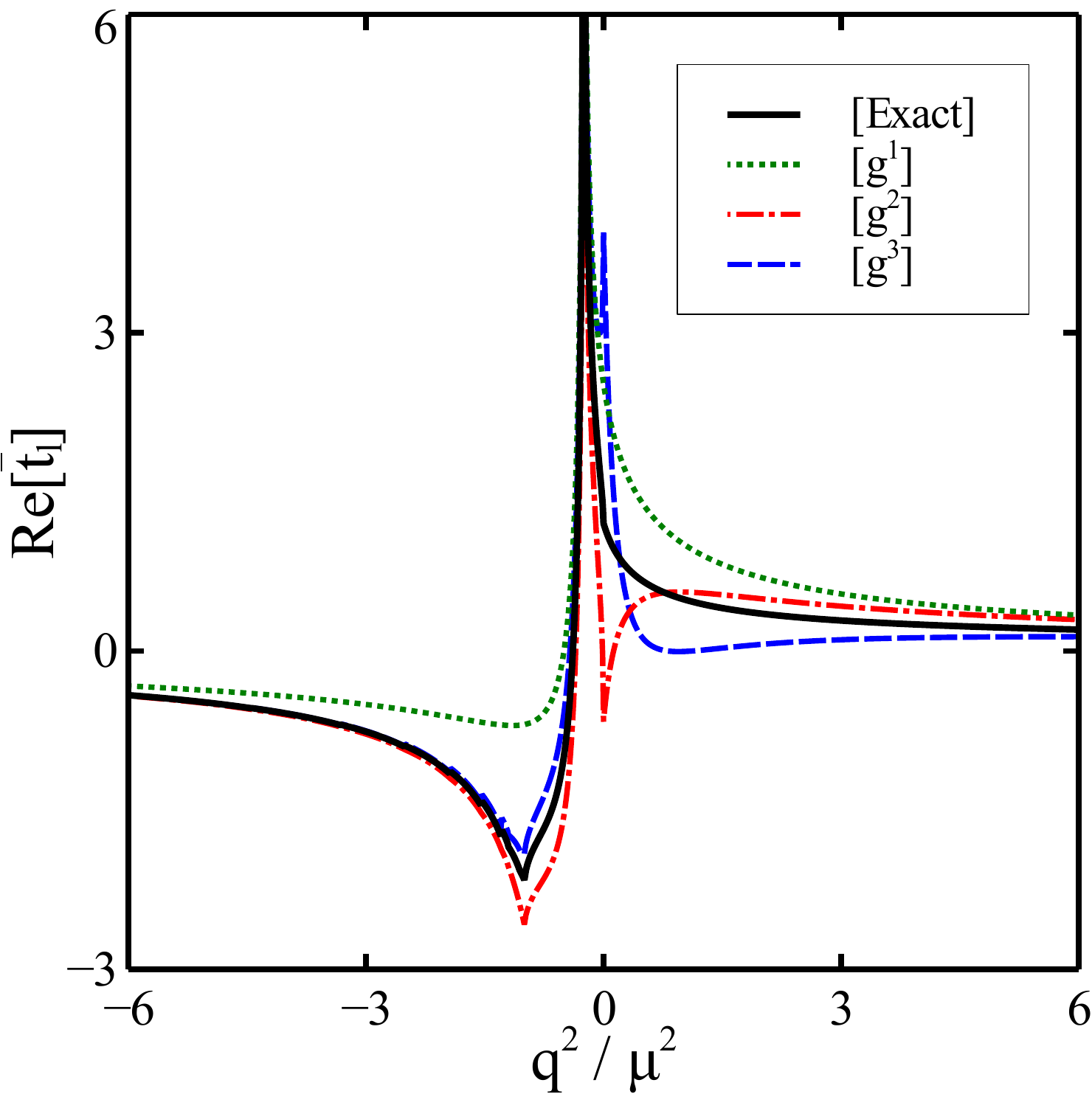}
  \includegraphics*[width=6cm,height=6cm]{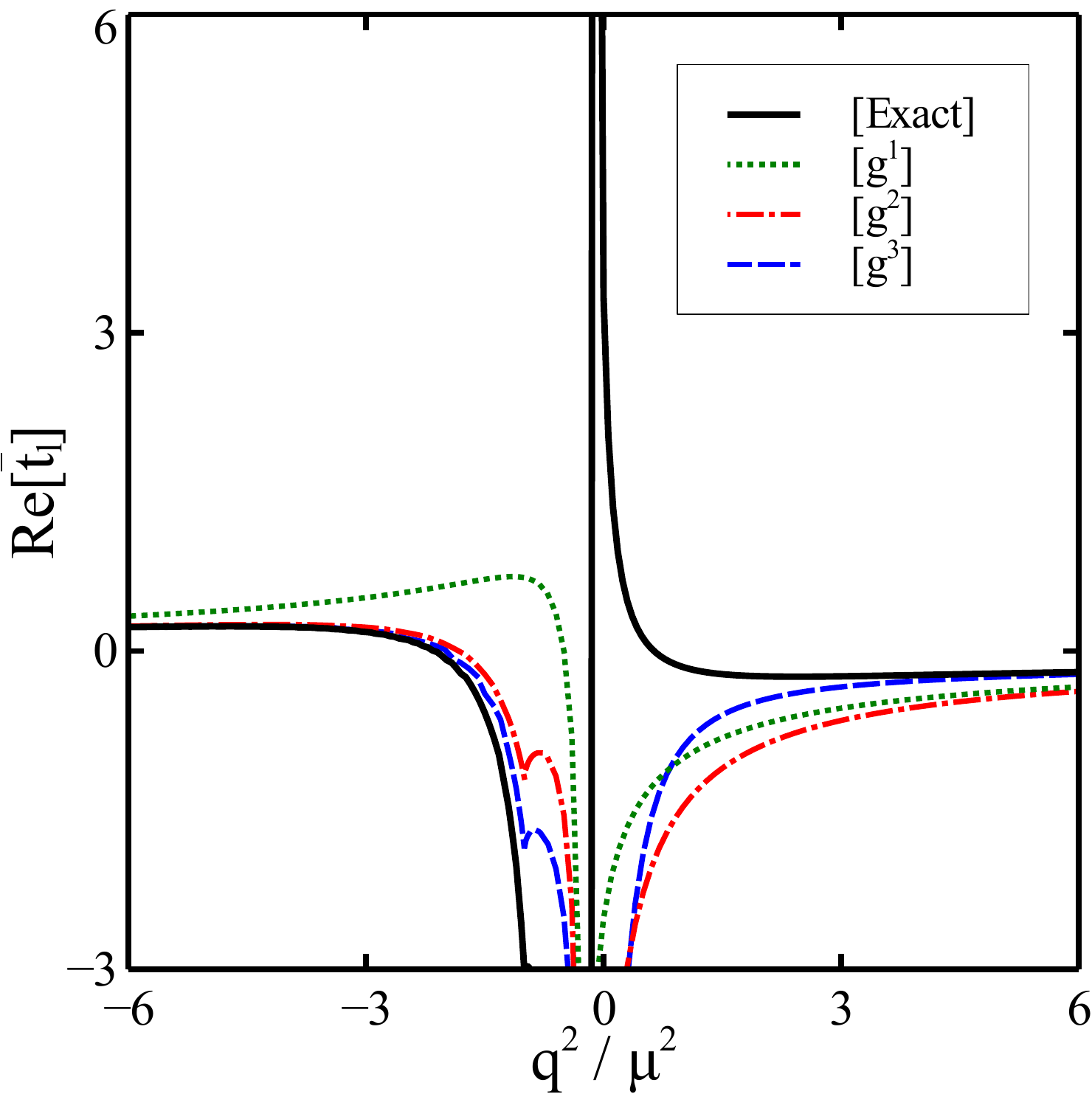}}
\caption{S-wave scattering amplitude for $|g|=\frac{3\,g_c}{2}$.  }
\label{fig:Born}
\end{figure}

We confirm the convergence properties by an explicit computation of the Born series to third order. In Fig. \ref{fig:Born}
the scaled on-shell amplitude
\begin{eqnarray}
\bar t_l \left(z \right) = 2\,m\,\mu \, \langle k' |\,t_l(q^2)| k\rangle \Big|_{k'=k=q}\qquad {\rm with } \qquad z = \frac{q^2}{\mu^2}\,,
\label{def-scaled-tl}
\end{eqnarray}
is shown for the particular choice $|g| = 3\,g_c/2$ in an s-wave with $l=0$.
The dimensionless amplitude $\bar t_l(z)$ is a function of the ratios $z= q^2/\mu^2$ and $g/g_c$ only.
While the left-hand panel in Fig. \ref{fig:Born} follows with positive and repulsive, the right-hand panel with negative and
attractive coupling constant. The figure illustrates convergence for large $|q^2|$. In contrast the near-threshold
region cannot be described by the Born series. Most striking is the bound state structure which arises for negative coupling
constant. Higher partial waves can be treated similarly to the $s$-wave,
but the speed of convergence for them is in general higher, the critical coupling constants are significantly larger then
the one for the $s$-wave.

How to construct a systematic approximation of the partial-wave scattering amplitude for the non-perturbative case with
$|g| > g_c$? While for the given Yukawa potential this is a rather academic question this is not the case for
realistic systems that can typically not be solved exactly. The idea is to construct an expansion that preserves
the causality and unitarity constraints of the scattering amplitude \cite{Gasparyan:2010xz}.

For our further consideration it is useful to introduce yet another dimensionless scattering amplitude. While the amplitude
(\ref{def-scaled-tl}) illustrates the scale dependencies of the system, it suffers from kinematical constraints at threshold.
The angular momentum barrier leads to
\begin{eqnarray}
\bar t_l(z) \simeq z^l\, \qquad  {\rm for} \qquad  z \simeq 0\,.
\label{}
\end{eqnarray}
We introduce an on-shell partial-wave scattering amplitude
\begin{eqnarray}
T_l(q^2) = -2\,m\,\sqrt{\Lambda^2 + q^2}\left( \frac{\Lambda^2 + q^2}{q^2} \right)^{l}\,\langle k' |\,t_l(q^2)| k\rangle \Big|_{k'=k=q} \,,
\label{}
\end{eqnarray}
with the scale parameter $\Lambda$. By construction the dimensionless partial-wave amplitudes are finite at threshold with $q^2 =0$.
Moreover, for any finite value of $\Lambda $ the amplitudes vanish with large momentum $q^2$ as
\begin{eqnarray}
T_l(q^2) \simeq - g\, \frac{m}{2\,\sqrt{q^2}} \,\log \frac{q^2}{\mu^2} \, \qquad {\rm for} \qquad q^2 \gg \Lambda^2 \,.
\label{asymptotics}
\end{eqnarray}
The unitarity constraint takes the simple form
\begin{eqnarray}
&& \Im \,T_l(q^2) = |T_l(q^2)|^2 \,\rho_l(q^2) \,, \qquad \quad
\rho_l(q^2) = \left(\frac{q^2}{\Lambda^2+q^2}\right)^{l + \frac{1}{2}}\,,
\label{def-unitarity}
\end{eqnarray}
illustrating that the parameter $\Lambda$ characterizes the scale at which the phase-space function $\rho_l(q^2)$ approaches its asymptotic value $1$. If the unitarity condition (\ref{def-unitarity}) is satisfied the partial-wave scattering amplitude
can be parameterized by the phase shift $\delta_l(q^2)$ with
\begin{eqnarray}
&& T_l (q^2) \,\rho_l(q^2) = \frac{1}{2\,i}\,\Big( e^{2\,i\,\delta_l (q^2)} -1 \Big)\,.
\label{def-phase-shift}
\end{eqnarray}

The analytic properties of the scattering amplitude generated by a Yukawa potential is well understood (see e.g. \cite{Taylor:1983}).
A key observation is the known analytic structure of the generalized potential $U_l(q^2)$ introduced by
\begin{eqnarray}
U_l(q^2)=T_l(q^2)-\int_{0}^{\infty}\frac{dq'^2}{\pi}\,\frac{q^2+\mu^2_M}{q'^2+\mu^2_M}\,
\frac{\rho(q'^2)}{q'^2-q^2-i\epsilon}\,|T_l(q'^2)|^2\,,
\label{def-potential}
\end{eqnarray}
where we insist on a matching scheme (see e.g. \cite{Mandelstam:1963,Gasparyan:2010xz}). At the matching scale $q^2 = -\mu_M^2$ the
generalized potential and scattering amplitude take identical values. While the scattering amplitudes $T_l(q^2)$ possess a
branch point at $q^2 =0$, reflecting the unitarity cut, the generalized potential is analytic at $q^2\geq 0$. The potential is characterized by left-hand cuts only. For a Yukawa potential it has branch points at $q^2 = -\Lambda^2$ and
$q^2 = - \frac{1}{4}\,(n\,\mu)^2$ with $n= 1,2,... $  \cite{Taylor:1983}.

A crucial step is to apply a perturbative expansion in powers of the coupling constant $g$ not to the scattering amplitude
but rather to the generalized potential $U_l(q^2)$. Such an expansion converges even for $|g|>g_c$ if the matching
point $\mu_M$ is chosen outside the nonperturbative domain of Fig.~\ref{fig:convergence}.
Previous but somewhat different studies along this line are \cite{Fivel:1960,Luming:1964,Collins:1968}.

We focus on s-wave scattering with $|g| = \frac{3}{2}\,g_c$ and choose a matching scale $\mu^2_M=10 \,\mu^2$,
that lies well below the boundary of the nonperturbative region.  The parameter $\Lambda$ will be varied.
For a given truncation of $U_l(q^2)$ we use (\ref{def-potential}) to reconstruct the corresponding approximate scattering amplitude. The non-linear integral equation  can be solved by means of the $N/D$ technique \cite{Chew:1960iv,Frye:1963}. The scattering amplitude is decomposed into the form
\begin{eqnarray}
T(q^2)=\frac{N(q^2)}{D(q^2)}\,,
\label{def-NoverD}
\end{eqnarray}
where the function $N(q^2)$ has only left-hand singularities, and $D(q^2)$ has only right-hand
singularities. These functions obey the following system of linear equations
\begin{eqnarray}
N(q^2)&=&U(q^2)+\int_{0}^\infty \frac{dq'^2}{\pi}\,\frac{q^2+\mu^2_M}{q'^2+\mu^2_M}\,
\frac{N(q'^2)\,\rho(q'^2)\,(U(q'^2)-U(q^2))}{q'^2-q^2}\,,
\nonumber\\
D(q^2)&=&1-\int_{0}^\infty \frac{dq'^2}{\pi}\,\frac{q^2+\mu^2_M}{q'^2+\mu^2_M}\,
\frac{N(q'^2)\,\rho(q'^2)}{q'^2-q^2-i\epsilon}\,.
\label{NoverD}
\end{eqnarray}

\begin{figure}[t]
\centerline{  \includegraphics*[width=6cm,height=6cm]{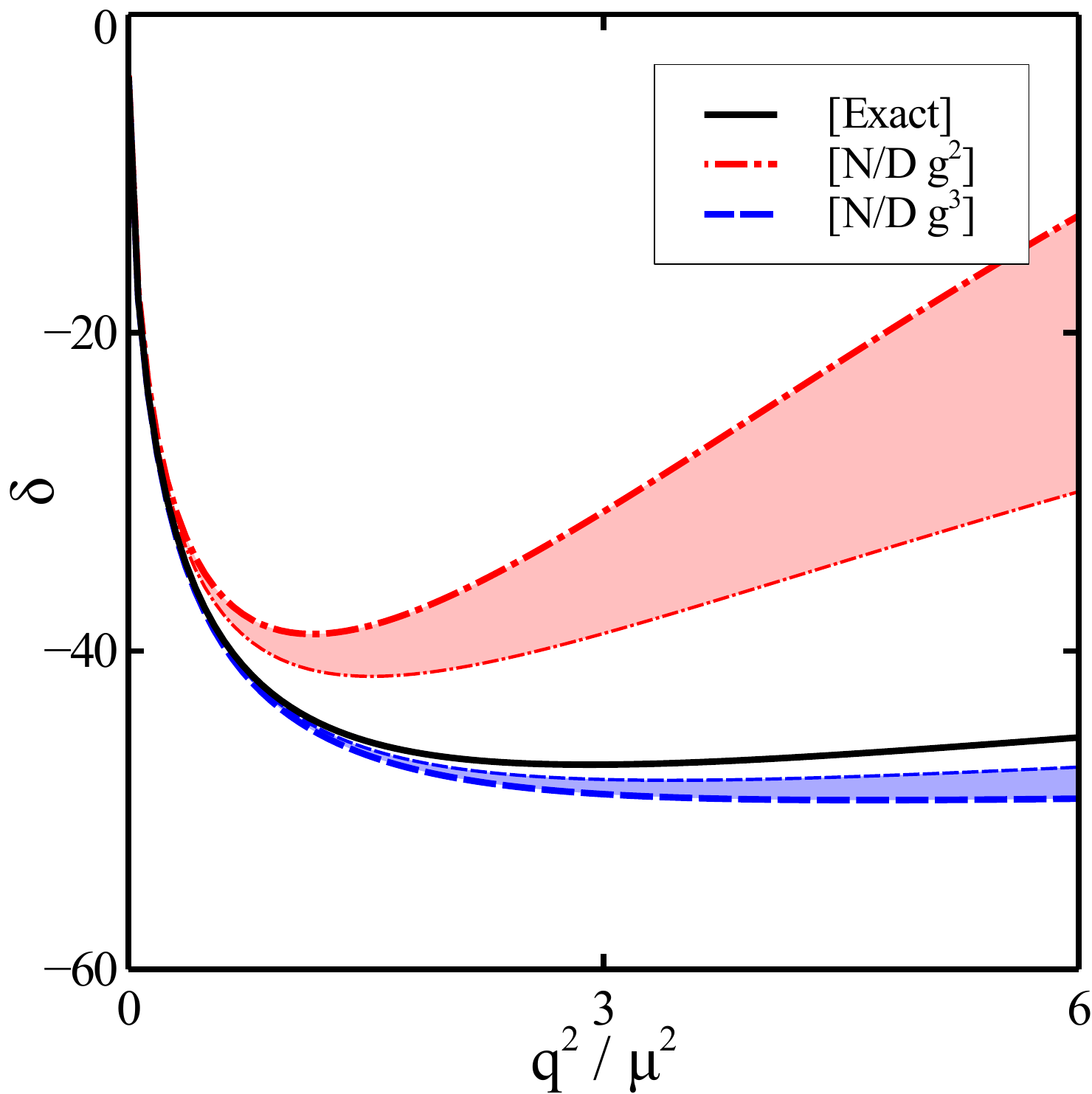}
  \includegraphics*[width=6cm,height=6cm]{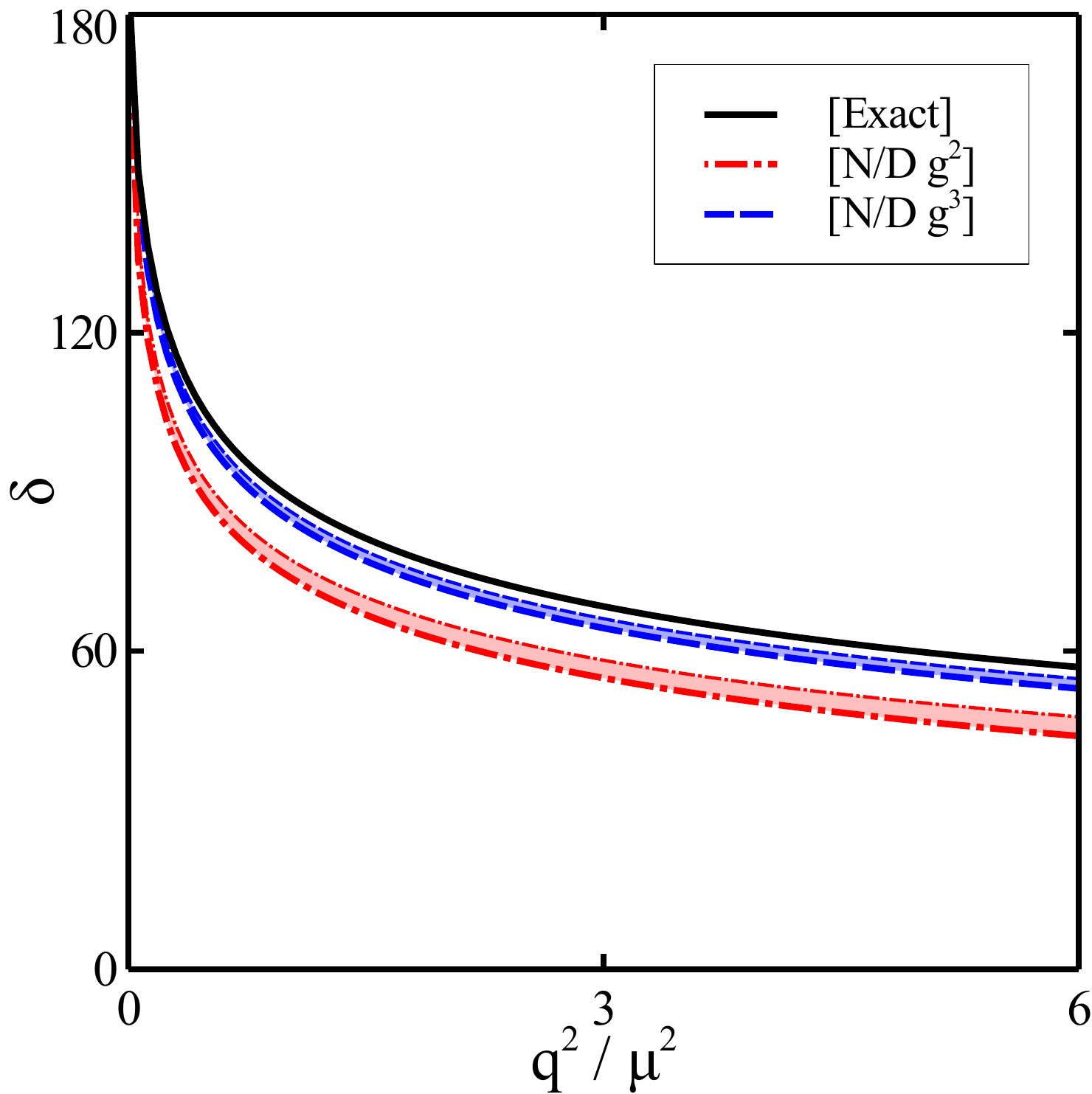}}
\caption{Scattering amplitude for $g=\frac{3}{2}\,g_c$ (left) and $g=-\frac{3}{2}\,g_c$ (right). The various bands are implied by a variation of parameter $\Lambda $ with
$3\,\mu < \Lambda < 9 \,\mu$.}
\label{fig:phases-1}
\end{figure}

In Fig.~\ref{fig:phases-1} we present the result of our numerical simulation of (\ref{NoverD}) for various truncations
of the generalized potential. We verified that indeed for all considered cases a solution of the non-linear integral
equation (\ref{def-potential}) was generated. In order to offer a more quantitative analysis we provide the scattering phase shift rather than the s-wave scattering amplitude. The figure demonstrates that the perturbative evaluation of the generalized
potential recovers the exact phase shift with increasing precision uniformly in energy. This is in contrast with the Born series,
which does not recover the scattering amplitude in the threshold region. Already a truncation of the generalized potential accurate
to cubic order reproduces the exact phase shift quite accurately.  The result is insensitive on the parameter $\Lambda$.
The spread implied by a variation $3\,\mu < \Lambda < 9 \,\mu$ is indicated by the bands. As expected the band width is decreasing
systematically as the order of the truncation increases.

For the attractive case there is a bound state in the system.
In the solution of (\ref{NoverD}) it appears as a zero in the $D$-function and therefore the scattering amplitude
develops a pole at $q^2= -q^2_B$. The binding energy is $E_B =q_B^2/(2\,m)$. In this case the solution of (\ref{NoverD}) does not constitute a solution of the nonlinear integral equation
(\ref{def-potential}). The truncated potential does not have a pole contribution. However, the solution of (\ref{NoverD})
is the solution of a related system where the pole term
\begin{eqnarray}
\frac{(q^2+\mu^2_M)\,g_B^2}{(q^2+2\,m\,E_B)\,(2\,m\,E_B-\mu^2_M)}
\label{def-add-pole-term}
\end{eqnarray}
is added to the approximated potential. It is straightforward to show that the set of corresponding N/D equations~(\ref{NoverD}) defines a scattering amplitude, which coincides with the one in the absence of (\ref{def-add-pole-term})
provided that the pole residuum $g_B$ is dialed properly.

Though the result of Fig.~\ref{fig:phases-1} is encouraging it does not yet permit a direct generalization to
realistic systems. In effective field theories, such as Chiral Perturbation Theory one can perform a perturbative
expansion only for small momenta. At first it appears impossible to arrive at a generalized potential that can be
trusted in the region needed to reconstruct the partial-wave scattering amplitude via the non-linear integral equation (\ref{def-potential}). The crucial observation is that the potential is required for $q^2>0$ only. This together with the
absence of branch points at $q^2>0$ in the potential permits the set up of a controlled approximation strategy for the
generalized potential.

We split the potential into an inside and outside part
\begin{eqnarray}
U(q^2)=U_{\rm inside}(q^2)+U_{\rm outside}(q^2)\,,
\label{}
\end{eqnarray}
where $U_{\rm inside}(q^2)$ contains the cuts, closest to the physical region and therefore have the largest effect on the energy
dependence of the generalized potential. Moreover, the inside part of the potential can be obtained with ease in particular
within effective field theories. The strength of an effective field theory is its capability to
compute threshold properties quite accurately and systematically. All contributions from the more distant cuts are subsumed in $U_{\rm outside}(q^2)$. Typically, the more distant cut structure are quite difficult to compute. In effective
field theories most frequently this is impossible.

In the case considered it is natural to include into $U_{\rm inside}(q^2)$ the cut from $-\mu^2<q^2<-\frac{\mu^2}{4}$ that is
determined exclusively by the first Born term. It takes the form
\begin{eqnarray}
U_{\rm inside}(q^2)=g\, m\, \int\limits_{-\mu^2}^{-\frac{\mu^2}{4}}\,\frac{dq'^2}{q'^2-q^2}\,\frac{\sqrt{q'^2+\Lambda^2}}{4\,q'^2}\,.
\end{eqnarray}
In turn the outside part of the potential is characterized by branch cuts in the interval $(-\infty,-\mu^2)$.

If the Taylor expansion of $U_{\rm outside}(q^2)$ around $q^2=0$ is known, we can use the method of conformal mapping to analytically extrapolate the generalized potential onto the whole positive real axis with $ q^2 \in (0, \infty)$. Even though in our case the Taylor series converges for   $|q^2|<\mu^2$ only, the knowledge of the Taylor coefficients permits an accurate reconstruction of the outside potential in the region required in (\ref{NoverD}). This is achieved in application of the transformation
\begin{eqnarray}
 \xi(q^2)=\frac{\mu-\sqrt{q^2+\mu^2}}{\mu+\sqrt{q^2+\mu^2}}\,,
\end{eqnarray}
which maps the cut complex plane onto the unit circle around $\xi(0) =0 $. The generalized potential can be reconstructed
with
\begin{eqnarray}
 U(q^2)=U_{\rm inside}(q^2)+\sum_{k=0}^\infty C_k \left[\xi(q^2)\right]^k\,.
\label{xi_expansion}
\end{eqnarray}
If the first $n$ coefficients in a Taylor expansion of $U_{\rm outside}(q^2)$
are known, one can compute $C_k$ for $k\le n$. By construction the expansion (\ref{xi_expansion})
converges in the cut complex plane.

\begin{figure}[t]
\centerline{  \includegraphics*[width=6cm,height=6cm]{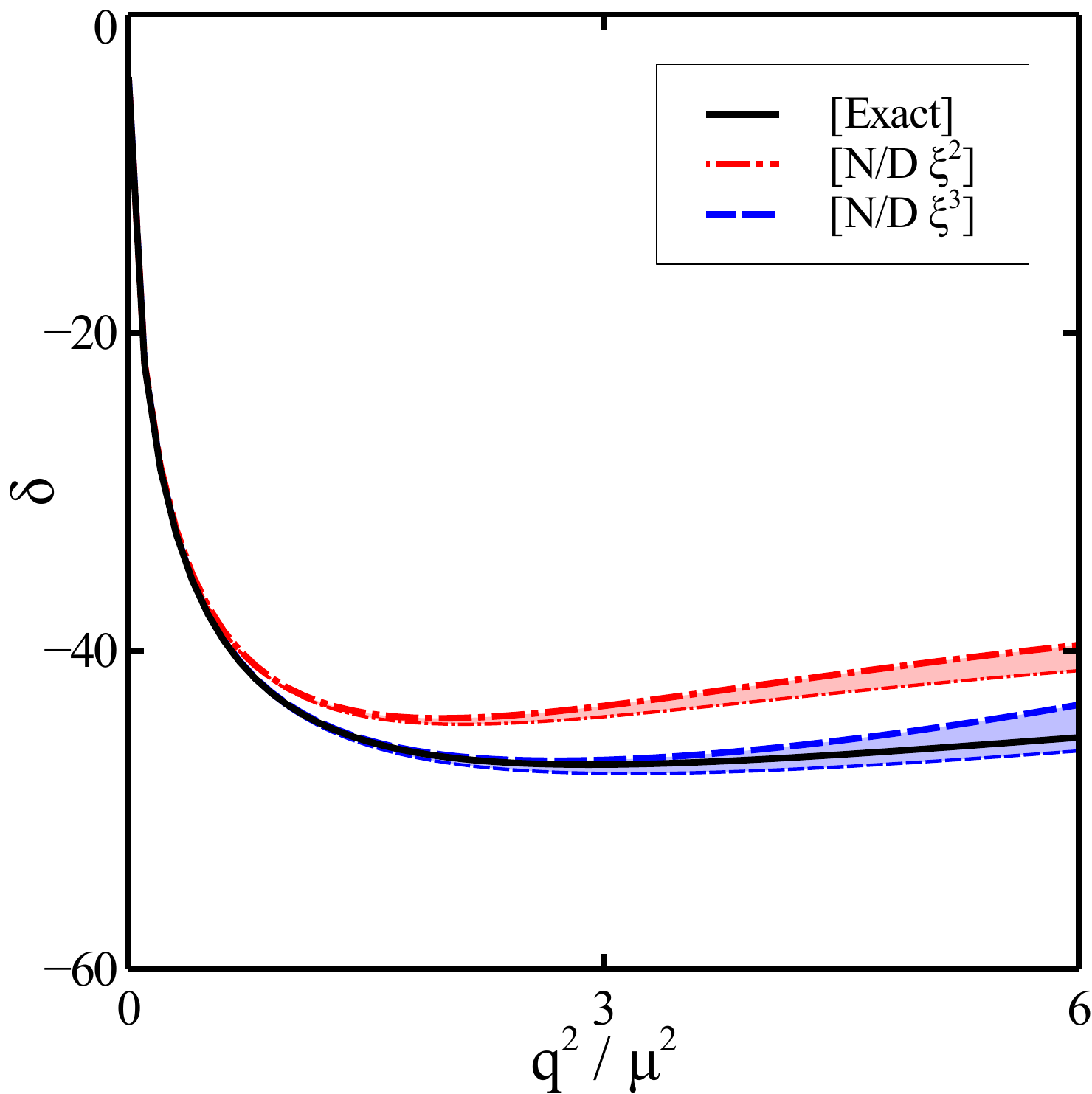}
  \includegraphics*[width=6cm,height=6cm]{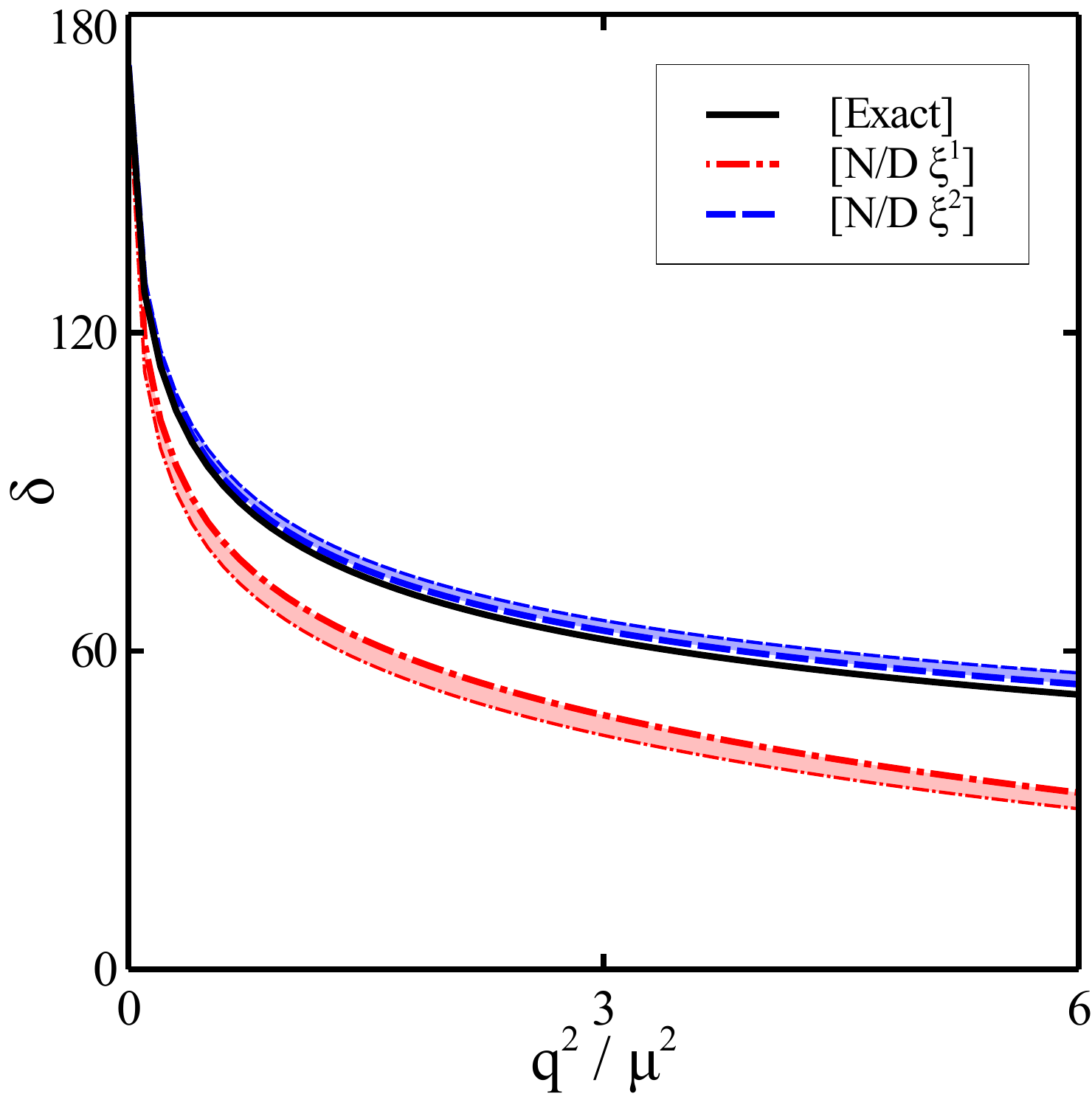}}
\caption{Scattering amplitude for $g=\frac{3}{2}\,g_c$ (left) and $g=-\frac{3}{2}\,g_c$ (right).
The various bands are implied by a variation of parameter $\Lambda $ with
$3\,\mu < \Lambda < 9 \,\mu$.}
\label{fig:xipansion}
\end{figure}

We studied the convergence properties of (\ref{xi_expansion}) for the examples
considered above. An expansion to the exact potential of the form (\ref{xi_expansion}) was performed. The results
are collected in Fig. \ref{fig:xipansion}. A rapid convergence is observed for the repulsive and attractive system.
An accurate reproduction of the exact phase shift is achieved considering the first three or four terms in
(\ref{xi_expansion}) only. Like in Fig. \ref{fig:phases-1} the sensitivity of our result on the
parameter $\Lambda$ is illustrated by bands. The band width decrease systematically with increasing order of the
expansion. For the sake of clarity  only two approximative results are confronted in Fig. \ref{fig:xipansion} with the exact
solid line.

With (\ref{xi_expansion}) we present an alternative expansion scheme for the generalized potential
that appears more practical in realistic applications. The coefficients $C_k$ may be computed
approximatively in perturbation theory or may be adjusted to empirical data directly.

\section{Expansion after renormalization }

The system described in the previous section shows some partial analogy with effective field theories
which may be constructed to be perturbative within the Mandelstam triangle. In contrast to the Yukawa toy model
in a realistic system the matching scale $\mu_M$ cannot be chosen arbitrarily low. Nonperturbative effects from
the $u$- or $t$-channel restrict the perturbative domain from below. The matching
should be performed within the Mandelstam triangle. In order to account for this requirement we set
$\mu_M=\mu$ in what follows. This particular choice implies a  matching right at the t-channel unitartity
branch point.

A further important issue is the intricate interplay of short and long-range forces. Typically the long-range forces
are rather week and perturbative but the short-range forces are strong and nonperturbative. For example in nucleon-nucleon scattering the central part of the pion-exchange potential has a coupling constant that is three times smaller than its critical value  \cite{Ericson:1988gk}. On the other hand there is often a strong short-range contribution, which can be responsible for the formation of bound states or resonances. The nature of short-range interactions is not resolved by effective field theories. They get renormalized
as to meet the physical conditions. We model this situation by considering an s-wave system defined by a superposition of two Yukawa potentials
\begin{eqnarray}
\langle k'|V_0 |k\rangle
&=&\langle k'|V_L|k\rangle+\langle k'|V_S|k\rangle\nonumber\\
&=&\frac{g_L}{4\,k'k}\,\log\frac{(k+k')^2+\mu_L^2}
{(k-k')^2+\mu_L^2}+
\frac{g_S}{4\,k'k}\,\log\frac{(k+k')^2+\mu_S^2}
{(k-k')^2+\mu_S^2},
\label{two_Yukawas}
\end{eqnarray}
where the second term is short range in the sense that $\mu_S \gg \mu_L$.

We renormalize the short-range force as to arrive at a closer correspondence to effective
field theories. In an initial step we treat the short-range part of the potential exactly but seek a perturbative treatment of
the long-range part. For this purpose it is useful
to introduce a compact matrix notation for (\ref{LSPW}). We write
\begin{eqnarray}
t&=&\big[ 1-(V_L+V_S)\,G\,\big]^{-1}\,\big[V_L+V_S\big]
\nonumber\\
&=&t_S+ (1+t_S \,G)\,V_L \sum_{n=0}^\infty \big[(G+G\, t_S \,G)\,V_L\big]^n\,(1+ G\,t_S)\,,
\label{series_long}
\end{eqnarray}
where $t_S= V_S+V_S \,G \,t_S$ is the full $t$-matrix in the presence of the short-range
potential alone. The crucial observation is that a contribution which involves $n$ long-range potentials $V_L$ and
$k$ short-range structures $t_S$ is characterized by its left-hand cut starting at
\begin{eqnarray}
q^2 = -\frac{1}{4}\,(n\,\mu_L+k\,\mu_S)^2 \,.
\label{}
\end{eqnarray}
Thus, in the limit as $\mu_S \to \infty$ the generalized potential takes the simple form
\begin{eqnarray}
U(q^2)=U_L(q^2)+ C\, \qquad {\rm with } \qquad  C = T(-\mu^2_M)-U_L(-\mu^2_M)\,,
\label{U_renormalized}
\end{eqnarray}
at least in the region, $q^2 >- \mu_M^2$, relevant for (\ref{NoverD}). The potential $U_L(q^2)$ is the one implied by the
long-range force only. It was studied in some detail in the previous section. The size of the coupling constant $g_S$ can be used to dial the value of the scattering amplitude at the matching point. This is nothing but the desired renormalization condition which
we will be insisting on in the following. Owing to this condition one may view our scheme as an analytic continuation of the
scattering amplitude from the matching point onwards into the physical region.

\begin{figure}[t]
\centerline{  \includegraphics*[width=6cm,height=6cm]{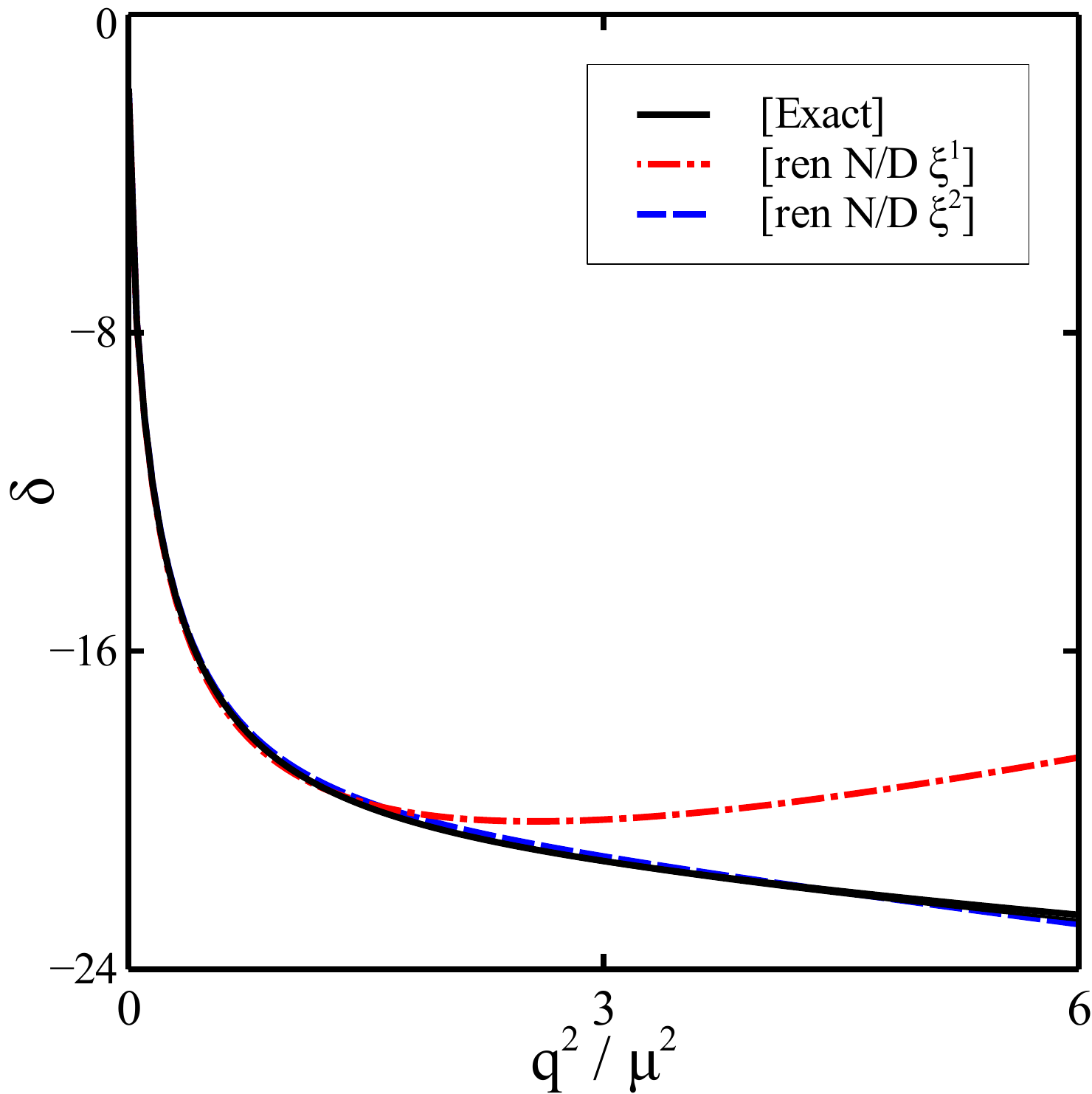}
  \includegraphics*[width=6cm,height=6cm]{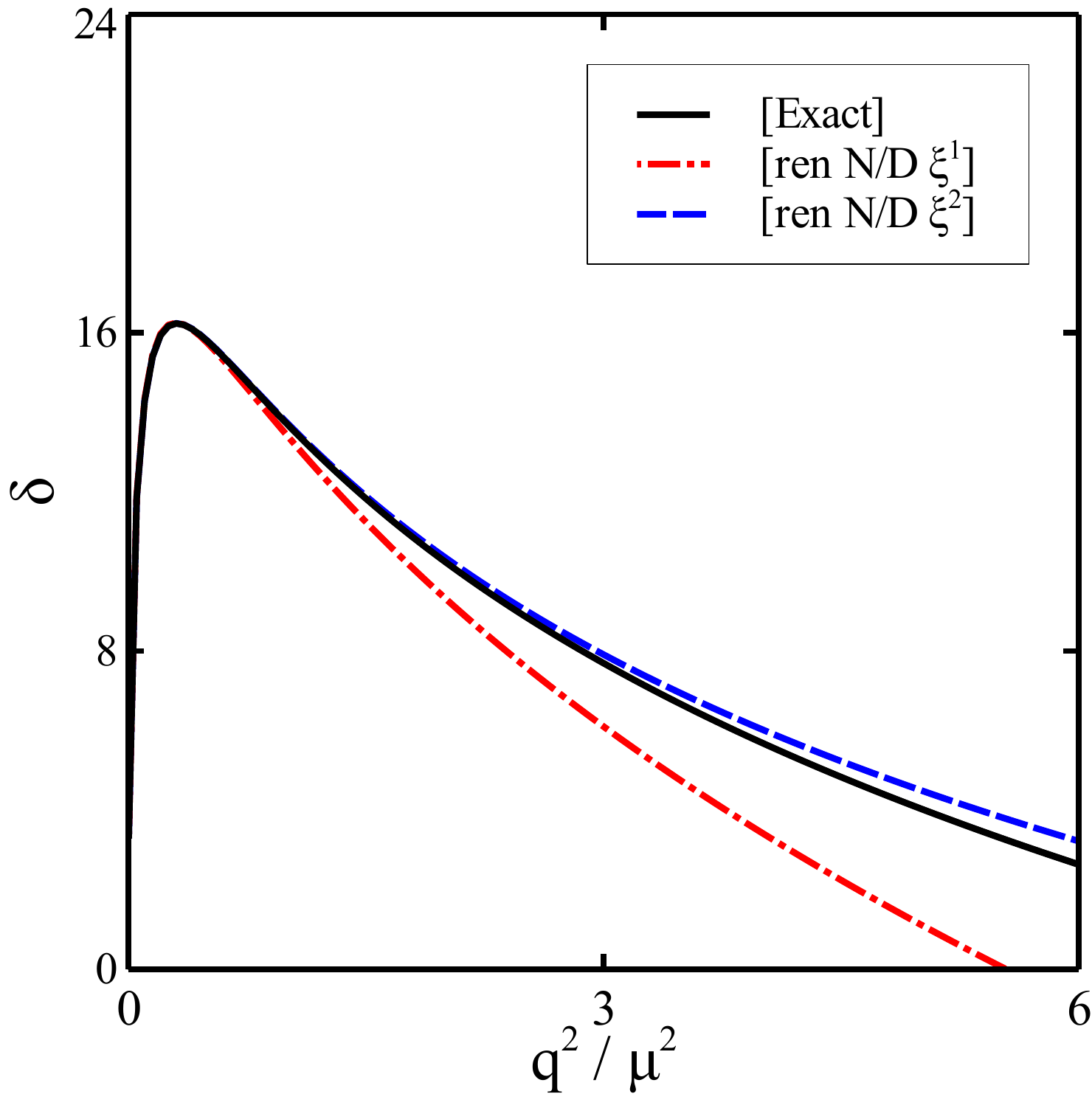}}
  \caption{Case study with repulsive short-range force. The l.h.p. and r.h.p. follow with
  repulsive and attractive long-range potential.
  }
\label{fig:renormalizationA}
\end{figure}

\begin{figure}[b]
\centerline{ \includegraphics*[width=6cm,height=6cm]{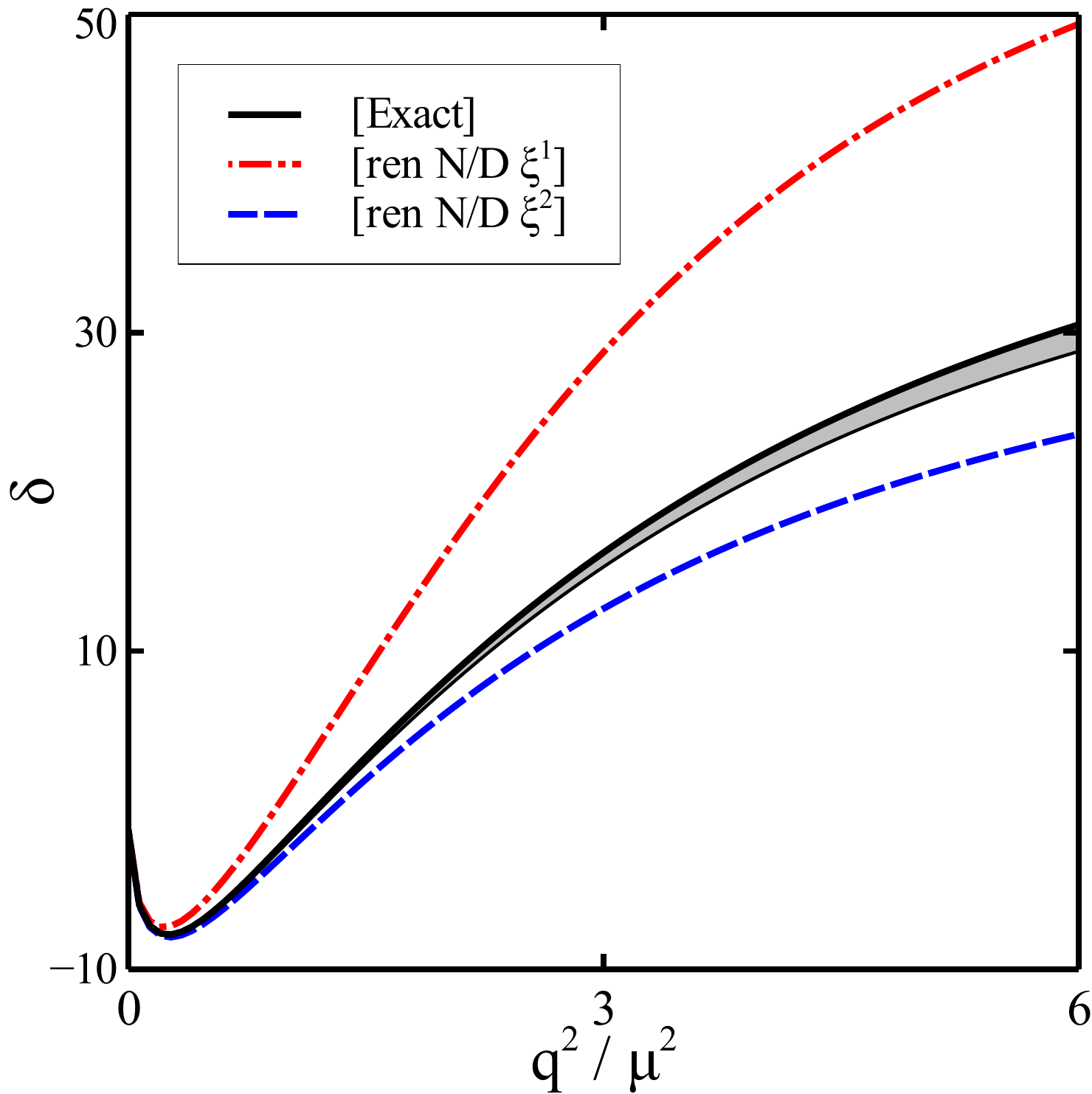}
  \includegraphics*[width=6cm,height=6cm]{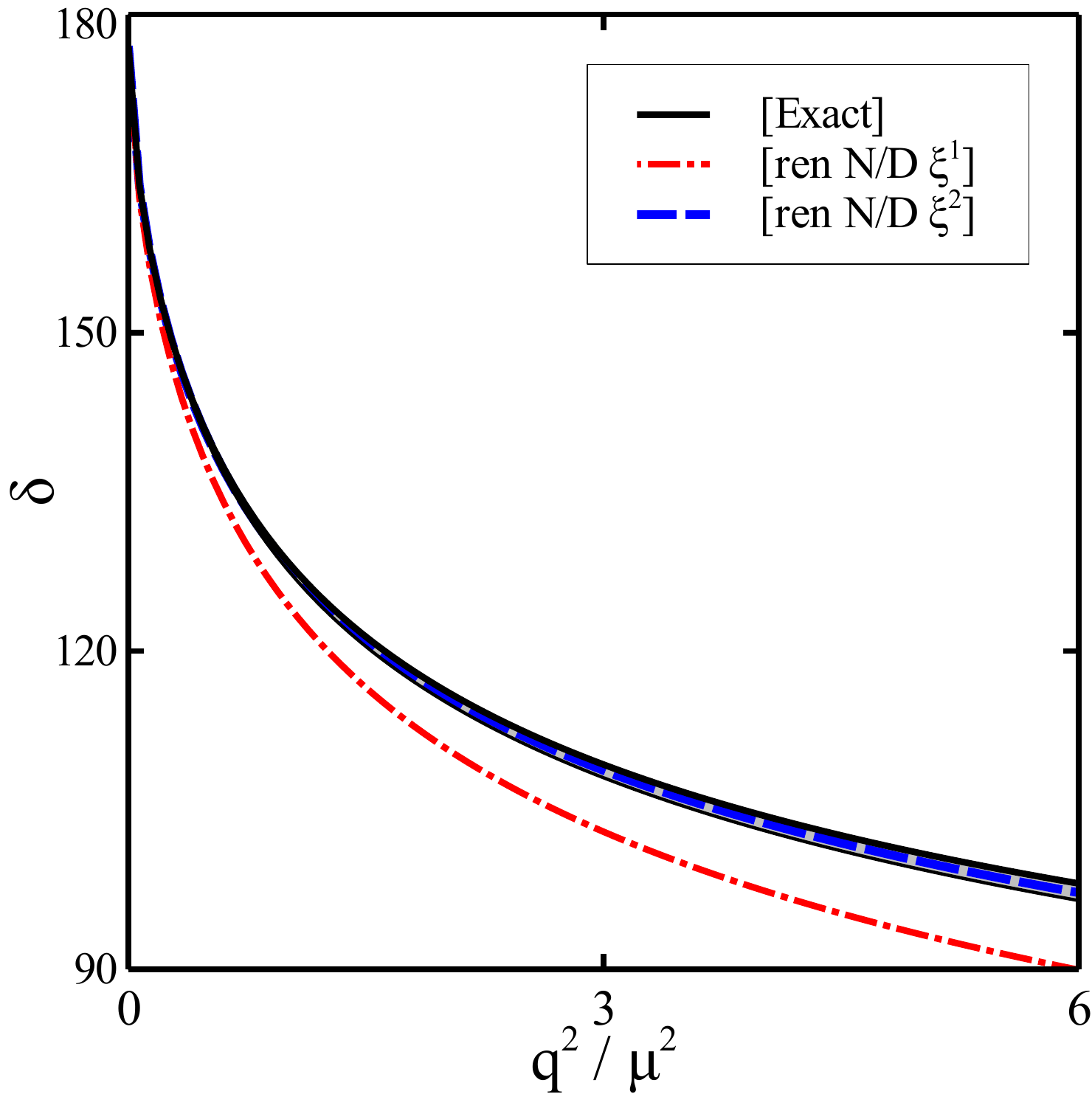}}
  \caption{Case study with attractive short-range force. The l.h.p. and r.h.p. follow with
  repulsive and attractive long-range potential.
  }
\label{fig:renormalizationB}
\end{figure}

\begin{figure}[t]
\parbox[c]{7.5cm}{\includegraphics*[width=6cm,height=6cm]{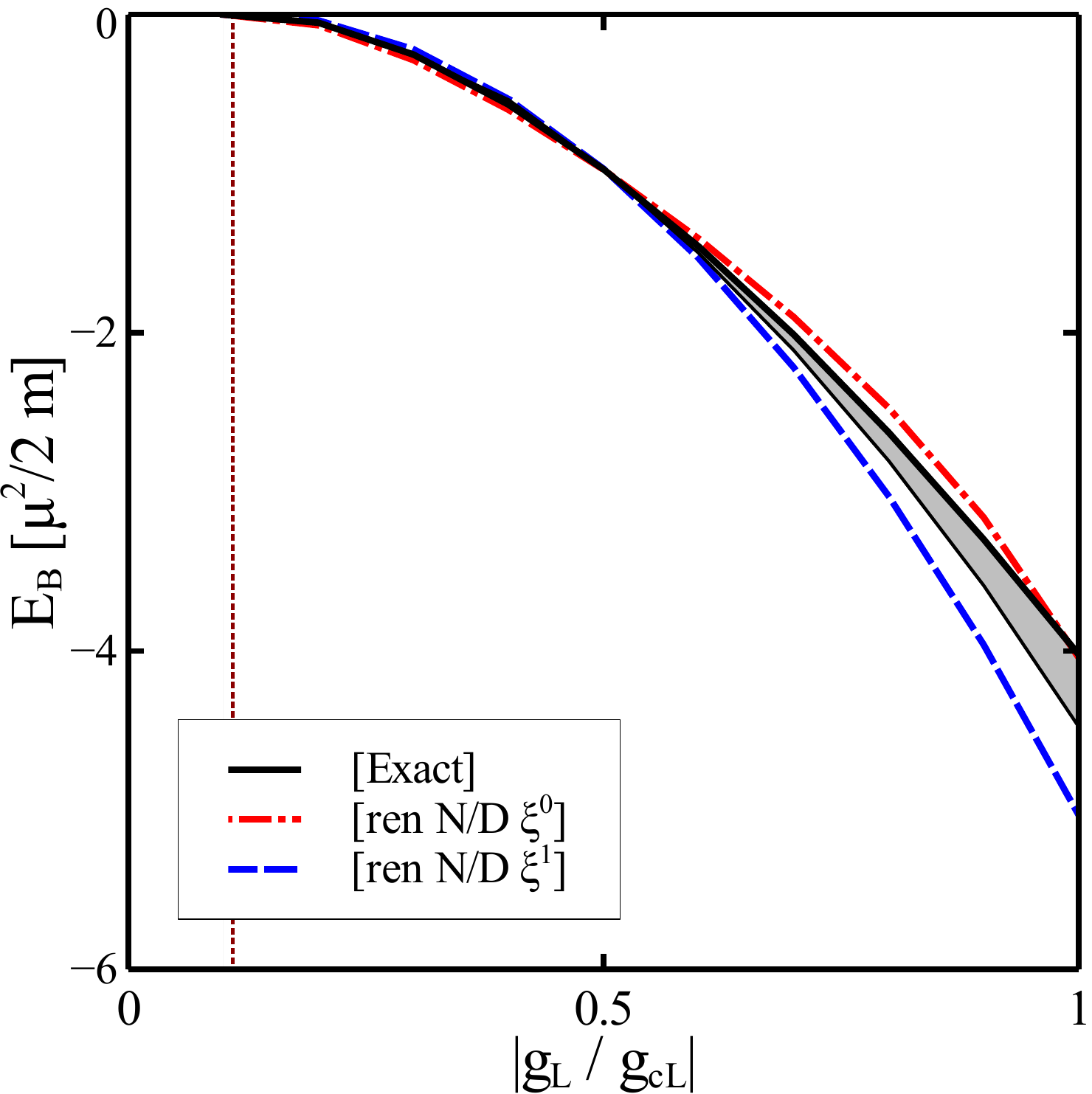}}
 \parbox[c]{5.0cm}{ \caption{Binding energy for the system with two attractive Yukawa interactions as a function of $g_L$.  }}
\label{fig:binding_energy}
\end{figure}

We consider the case of a repulsive and attractive short-range interaction with $\mu_S=12\,\mu_L $ for definiteness.
At the scale of the long-range potential the short-range potential can be viewed as pointlike. We set
\begin{eqnarray}
|g_S|=0.95\,g_{c,S} \simeq 0.80 \,\frac{\mu_S}{m}\,,
\label{def-gS}
\end{eqnarray}
so that the interaction is rather strong, but a bound state is not yet formed for the  attractive case.
For the long-range part of the potential we take the coupling constants $g= \pm \frac{g_c}{2}$.
The influence of the left-hand cut implied by the short-range potential can be further weakened by a correlation of the
parameter $\Lambda$ and the Yukawa mass $\mu_S$. For
\begin{eqnarray}
\Lambda = \frac{1}{2}\,\mu_S\,,
\label{de-correlation}
\end{eqnarray}
a smooth spectral weight at the leading branch
point with $q^2 = -\mu_S^2/4$ arises. Any different choice leads to a logarithmic divergence at that branch point. In the following
we use (\ref{de-correlation}) as it implies a significantly more rapid realization of (\ref{U_renormalized}) as $\mu_S \to \infty$.

Given the parameter set we compute the phase shift for the four different scenarios.
In Fig. \ref{fig:renormalizationA} and Fig. \ref{fig:renormalizationB} the exact results are confronted with the
approximation implied by (\ref{U_renormalized}). The shaded area in the figures illustrate the accuracy
of (\ref{U_renormalized}). While for the repulsive short-range force the thick and thin solid lines are
almost indistinguishable, we observe small discrepancies for the attractive short-range force. The thin lines correspond to the approximate results with (\ref{U_renormalized}). The additional dashed and dashed-dotted lines follow with the long-range potential $U_L(q^2)$ in (\ref{U_renormalized}) truncated according to (\ref{xi_expansion}). In all cases already the second order provides a reasonable approximation, where the convergence is more rapid for the repulsive short-range force.

We further scrutinize the quality of our approximation scheme. Fig.~7 shows the dependence of the binding energy on $g_L< -0.11 g_{c L}$
of the bound state that is formed for the attractive short-range potential and a sufficiently strong long-range force.
Shown is the exact binding energy together with the approximations defined by (\ref{U_renormalized}) and (\ref{xi_expansion}). The
solid lines are as in the two previous figures.  The determination of the binding energy is
quite accurate already at zeroth order in the conformal mapping for a wide range of couplings constants.

We summarize that the constraints set by causality and unitarity can be used to arrive at a quite effective expansion
scheme that is suitable for applications in effective field theories \cite{Gasparyan:2010xz}.

\bibliography{1}
\bibliographystyle{elsart-num}

\end{document}